\documentclass[%
 preprint,
 nopreprintnumbers,
showpacs,
nobibnotes,
 amsmath,amssymb,
]{revtex4-1}

\usepackage{graphicx}
\usepackage{dcolumn}
\usepackage{bm}
\usepackage{epstopdf}

\newcommand{\angstrom}{\mbox{\normalfont\AA}}
\usepackage{xcolor}
\newcommand{\fig}[1]{Fig.~\ref{fig:#1}}
\newcommand{\eq}[1]{Eq.~(\ref{eq:#1})}
\newcommand{\tab}[1]{Table~\ref{tab:#1}}

\usepackage{float}

\begin{document}

\title{Shadow wave function with a symmetric kernel}

\author{V.~Zampronio}
\thanks{These two authors contributed equally to this work}
\email{viniciuz@ifi.unicamp.br - vpedroso@ifi.unicamp.br}
\author{V.~Z.~Pedroso\normalfont\textsuperscript{*}}%
\author{S.~A.~Vitiello}
\email{vitiello@ifi.unicamp.br}
\affiliation{%
Instituto de F\'{i}sica Gleb Wataghin, Universidade Estadual de Campinas,
UNICAMP, 13083-859, Campinas, S\~{a}o Paulo, Brazil
}%




\date{\today}

\begin{abstract}
A shadow wave function with an explicit symmetric kernel is introduced.
As a consequence
the atoms exchange in the system is enhanced.
Basic properties of this class of trial functions are kept and
quantities it can describe are easily estimated. The effectiveness of this approach is
analized by computing properties of interest in a system formed from
$^4$He atoms.
\end{abstract}             

\maketitle


\section{\label{sec:intro}Introduction}

Quantum matter that shows effects at the macroscopic level has attracted
attention of physicists for many decades.
One of the most studied systems presenting this behavior is
formed from helium atoms.
The richness of phenomena
observed in both the liquid and solid phases of helium justify an
interest that persists until today.

Variational theories are important tools in the investigation of quantum
many-body systems. They are able to give physical insight in the
processes of interest based on our physical intuition.
The variational investigation of systems
formed from helium have also a long history.
Soon it was recognized that
strong interactions between the atoms at short-range distances needed
to be taken in an explicit way.
First  successful
Monte Carlo calculations in the liquid phase were done using a trial
function of the Bijl-Dingle-Jastrow form \cite{mcm65,*sch67}.
One of the further improvements at the level of two-body correlations where made by
introducing a basis set to optimize pair functions \cite{vit99}.
Beyond the pair product wave functions, the introduction of explicit
three-body terms were able to improve the overall description of the
helium systems \cite{sch80,*sch81}.
Properties of the solid phase of these highly anharmonic crystals where
initially computed by explicitly introducing an \textsl{a priori} lattice
\cite{han68},
as suggested by Nosanow.
Although good agreement with experiment was obtained with this approach,
it was at a cost of spoiling translational invariance and the Bose
character of the wave function.
In a relative recent effort,
a variational ansatz have restored the
Bose symmetry in the Nosanow-Jastrow
description of $^4$He and presented interesting
results for the solid-liquid phase transition of this quantum system
\cite{lut14}.

Apart from these variational ans\"atze,
a different class
of trial functions, the shadow functions,
were introduced
long ago \cite{vit88,mac94}.
Its ideas are widely employed in the investigation of a variety of systems
\cite{cal15,*cal14,*gal14,*pes10}.
Maybe the simplest motivation of this class of variational functions is
to think about the auxiliary variables, used in their definition, as the
center of mass of polymers that represent each atom in Feynman's
path-integral approach in imaginary time.
The shadow wave functions are
translational invariant and Bose symmetric functions.
Although it
implicitly correlates particles up to the number of bodies present in
the system, the functional form of these correlations are unknown.
This work is an attempt to improve these correlations by explicitly
symmetrizing a kernel that couples the atoms and the auxiliary variables. This is a way of explicitly
emulate the cross-link between the polymers in
Feynman's path-integrals.
An immediate benefit from this approach is
the possibility of
estimate the momentum distribution function as done easily by McMillan
\cite{mcm65}, in much simple calculation than
previously done when considering shadow wave functions\cite{vit90}.
With the aim of testing the consequences of an explicit symmetric kernel
in a shadow function we
investigated several properties of
the systems formed from
$^4$He atoms.

We have organized this work as follow.
In section \ref{sec:psissk} we introduce our shadow wave function with a symmetric kernel. 
The methods used in our calculations are presented in section \ref{sec:vmc}.
Results obtained for the variational energies, melting and freezing densities
and radial distribution functions are presented in section \ref{sec:results}. In this section
we also make a careful discussion of the condensate fraction associated with our
trial function. We show that the computation of this important quantity needs special attention.
The last section is devoted to final comments.

\section{\label{sec:psissk}A shadow wavefunction with a symmetric kernel}

The simplest
Hamiltonian used to describe a system of $N$ atoms of $^4$He is written as

\begin{equation}
\label{ham}
 H = -\frac{\hbar^2}{2m} \sum_{i=1}^N \nabla_i^2 + \sum_{i<j}^N V(r_{ij}),
\end{equation}

\noindent  where $m$ is the $^4$He mass, $r_{ij}$ is the distance between atoms $i$ and $j$, and $V$ is an inter-atomic pairwise potential.
In this work we use the He-He inter-atomic potential HFD-B3-FI1 as proposed by Aziz and co-workers\cite{azi95}.  

Our trial wave function
is constructed by the integration of auxiliary variables 
$S = \{{\bf s}_1, {\bf s}_2, \ldots, {\bf s}_N\}$ in the whole
space

\begin{equation}
\Psi_{SSK}(R) = \psi_a(R)\int d^3S \ \Xi_{SK}(R,S), 
\end{equation}

\begin{equation}
 \Xi_{SK}(R,S) = \Theta_{SK}(R,S)\psi_s(S),
\end{equation}

\noindent where $R = \{{\bf r}_1, {\bf r}_2, \ldots, {\bf r}_N\}$ is the
set of the atomic coordinates in the
configuration space.
The kernel $\Theta_{SK}(R,S)$, unlike shadow wave functions
forms earlier proposed, is symmetric under the exchange of atoms and
bounds each auxiliary variable with all atoms by a product of a sum of
Gaussian functions,

\begin{equation}
 \Theta_{SK}(R,S) = \prod_{j=1}^N\sum_{i=1}^N e^{-C|{\bf r}_i - {\bf s}_j|^2} ,
\end{equation}

\noindent where $C$ is a variational parameter.
This form
of $\Theta_{SK}$
was devised by
Cazorla {\it et al.} \cite{caz09} for a
symmetrization of a one-body  Nosanow factor.
An additional motivation for choosing this symmetric kernel is that it might
improve the exploration of the configuration space by explicitly
connecting all the atoms to all auxiliary
variables.

The functions $\psi_a(R)$ and $\psi_s(S)$ are
product of two-body factors
of the Jastrow form.
The function $\psi_a(R)$
correlates the atoms

\begin{equation}
 \psi_a(R) = \prod_{i<j}^N e^{-\frac{1}{2}u(r_{ij})},
\end{equation}

\noindent where $u(r)$ is a pseudo-potential of the McMillan form\cite{mcm65} with a variational parameter $b_a$,

\begin{equation}
 u(r) = \left( \frac{b_a}{r} \right)^5.
\end{equation}
The auxiliary variables are correlated by

\begin{equation}
 \psi_s(S) = \prod_{i<j} e^{-w(s_{ij})},
\end{equation}

\noindent most of our calculations were made with 
$ w(s) = \beta V(\alpha s)$,
the Aziz two-body inter-atomic potential
rescaled in its amplitude and distance by
variational parameters $\beta$ and $\alpha$.
For comparison,
at the equilibrium density in the liquid phase,
we have also considered a
pseudo-potential
of the McMillan
form, $w(s)=\left(b_s/s\right)^9$,
with a variational parameter $b_s$.

\section{\label{sec:vmc}The variational Monte Carlo calculations}

In the variational Monte Carlo (VMC) method the
trial energy can be written as
\begin{equation}
E_V = \int dR dS dS' {\cal P}(R,S,S') E_L(R,S) ,
\end{equation}
where
$E_L$ is the local energy,

\begin{equation}
E_L(R,S) = \frac{H \ \psi_a(R)\Theta_{SK}(R,S)}{\psi_a(R)\Theta_{SK}(R,S)}.
\end{equation}
This quantity can also be computed by the set $\{R,S'\}$.
The probability density function ${\cal P}(R,S,S')$,
of the
configurations in our simulations  is
given by the set of atomic coordinates and
two different sets of auxiliary variables,

\begin{equation}\label{eq:probability}
{\cal P}(R,S,S') =
\frac{\psi_a^2(R)\Xi_{SK}(R,S)\Xi_{SK}(R,S')}{\int
d^3 R' \ \Psi_{SSK}^2(R')}.
\end{equation}
A second set of auxiliary variables
is needed because we perform a simultaneous integration on the variables
$\{R,S,S'\}$ and the square of the wave function needs to be considered.

We compute the variational energy as averages over the sampled configurations

\begin{equation}\label{expected_value}
E_V  = \frac{1}{2}\left< E_L(R,S) + E_L(R,S')\right>,
\end{equation}

\noindent because this is more efficient, it will reduce the variance for a given
computer time. For all properties the sets S and S' are equivalents.

The sample was made using the Metropolis algorithm \cite{met53}. 
The configuration of the atoms, $R$, are sampled for fixed values of $S$
and $S'$. Each set of the auxiliary variables in its turn are sampled with
the $R$ configuration fixed. We may note that $S$ and $S'$ could naturally
be sampled in parallel.
Because of the particular form of our trial function, it is more
advantageous to attempt moves where all particles are considered at once.
To this aim, for the atoms we use
the pseudoforce $F_a$

\begin{equation}\label{quantumforce}
\begin{aligned}
& \ F_a(R,S,S') = \nabla_R\text{ln} [\psi_a^2(R) \  \\
& \ \times \ \Theta_{SK}(R,S)\Theta_{SK}(R,S') ].
\end{aligned}
\end{equation}

\noindent Moves of the atoms are proposed according to the expression

\begin{equation}\label{eq:moves_eq}
R_p = R + \sqrt{2D\tau_a}g + D\tau_a F_a ,
\end{equation}

\noindent where $D = \hbar^2/2m$, $g$ is a matrix of normal Gaussian random variables, 
and $\tau_a$ is a calculation parameter.
Moves are accepted with a probability given by

\begin{equation}
\begin{aligned}
& \ q_a(R,R_p)=\frac{\psi_a^2(R_p)}{\psi_a^2(R)} \times \\
& \ \times \frac{\Theta_{SK}(R_p,S)\Theta_{SK}(R_p,S')}{\Theta_{SK}(R,S)\Theta_{SK}(R,S')}\frac{T(R_p,R)}{T(R,R_p)},
\end{aligned}
\end{equation}

\noindent where $T$ is a transition matrix,

\begin{equation}\label{eq:tmatrix}
 T(R,R_p) = (4\pi D\tau_a)^{-\frac{3N}{2}}\text{e}^{-\frac{(R_p - R -D\tau_a F_a(R))^2}{4D\tau_a}}.
\end{equation}

For the shadow particles, moves are proposed in a similar way to \eq{moves_eq}, using
either $F_s$ or $F_{s'}$ with a parameter 
$\tau_s$. For the shadow particles $S$, the pseudoforce $F_s$ is computed through

\begin{equation}\label{quantumforce} F_s(R,S,S') = \nabla_S
\text{ln}[\Theta_{SK}(R,S)\psi_s(S)], 
\end{equation}
and an equivalent expression for the $S'$ particles.
The shadow particles moves are accepted with the probability

\begin{equation}
q_s(S,S_p) =\frac{\Theta_{SK}(R,S_p)\psi_s(S_p)}{\Theta_{SK}(R,S)\psi_s(S)}\frac{T(S_p,S)}{T(S,S_p)}.
\end{equation}

\noindent Similar expressions of \eq{tmatrix} for $S$ and $S'$ are employed when attempts are made
to change those variables, in those expressions $\tau_s$ and $F_s$ or
$F_{s'}$ are used instead of $\tau_a$ and $F_a$.

We have estimate the
total energy of a system made from $^4$He atoms
at some densities by minimization of the trial energy with respct to the variational parameters. 
Equations of state for the liquid and solid phases as a function of
the density were determined by fitting the coefficients of
a third degree polynomial to the
obtained total energies per particle

\begin{equation}\label{eq:eos}
 \frac{E}{N} = A + B\left(\frac{\rho - \rho_0}{\rho_0}\right)^2 + C\left(\frac{\rho-\rho_0}{\rho_0}\right)^3,
\end{equation}

\noindent where $\rho_0, A, B, C$ are fitting parameters.
At the liquid phase, it's easy to see that $\rho_0$ represents the
density of equilibrium at zero pressure. For the
solid phase this parameter does not have a particular meaning.

Once the variational minimization of the energies
as a function of de density was done, it is
interesting to investigate how the obtained trial functions
describe properties that do not satisfy a variational principle.
From the equations of state of the liquid and solid phases
we can easily obtain the freezing and melting densities, $\rho_f$ and
$\rho_m$, at $T = 0 \ \text{K}$ using the double tangent Maxwell construction
that consists in solving the following equations,

\begin{equation}\label{eq:mft}
\left\{ \begin{aligned} 
& \ \rho_f^2\left(\frac{\partial E}{\partial \rho}\right)_{\rho = \rho_f} = \rho_m^2\left(\frac{\partial E}{\partial \rho}\right)_{\rho = \rho_m}\\
& \ E_f - E_m = \rho_f^2\left(\frac{1}{\rho_m} - \frac{1}{\rho_f}\right)\left(\frac{\partial E}{\partial \rho}\right)_{\rho = \rho_f},
\end{aligned} \right.
\end{equation}

\noindent where we have used the notation
$E_{(.)}=E(\rho_{(.)})$.

The
condensate fraction  is another property
of interest that can be obtained from the
off-diagonal matrix element of the 
one-body density matrix

\begin{equation}\label{eq:density_matrix1}
  \rho_1({\bf r})
= N
\frac{\int d R \ \Psi(R')
\Psi(R)}
{\int d^3 R \ \Psi^2(R)},
\end{equation}

\noindent where $R' \equiv \{{\bf r}_1 + {\bf r}, {\bf r}_2, \ldots, {\bf r}_N\}$. For an homogeneous and isotropic system $\rho_1$ can depend only on the
magnitude of the displacement vector ${\bf r}$ and $\rho_1({\bf r}) = \rho_1(r)$. 

For a shadow wave function $\rho_1(r)$
can be expressed as

\begin{equation}\label{eq:density_matrix}
 \rho_1(r) = \left< \frac{\psi_a(R')\Theta_{SK}(R',S)}{\psi_a(R)\Theta_{SK}(R,S)}\right>.
\end{equation}
The symmetrical kernel we have implemented in our trial function
allows its evaluation with the configurations sampled from
the probability ${\cal P}(R,S,S')$ of \eq{probability}.
Previously\cite{mac94,vit90}, only if the integrand of the single-particle density matrix, of Eq.(\ref{eq:density_matrix1}), was
sampled it was possible to estimate $\rho_1(r)$
within Monte Carlo
calculations of feasible duration. This happened because of the Gaussian
coupling between the atoms and the shadows would lead $\rho_1(r)
\rightarrow 0 $ for large values of $r$.
Since the ergodicity of the sampling of these two
probability densities may vary, we have considered both
methods of computing $\rho_1$ to compare their results.

In the standard way\cite{mcm65}, given by Eq.(\ref{eq:density_matrix}), of computing $\rho_1$ an histogram is
constructed with bar width $\Delta r$ small enough to give a good
representation of $\rho_1(r)$.
For each
configuration we randomly choose an atom
in position ${\bf r}_i$, it is displaced to a random position ${\bf
r}_i'$ and
the distance $r = |{\bf r}_i - {\bf r}_i'|$
under periodic boundary conditions is evaluated. In the respective bin of
this distance, the ratio of \eq{density_matrix} is then
accumulated. With
this procedure we obtain an estimate of
$4\pi r^2 \Delta r  \rho_1(r)/N$.
Finally the fraction of atoms in the zero-momentum state can be obtained
 as\cite{pen56}

\begin{equation}
n_0 = \lim_{r\rightarrow \infty} \frac{\rho_1(r)}{\rho}.
\end{equation}

The second way we have considered of calculating $\rho_1(r)$ is by sampling the probability
density function associated to configurations of the off-diagonal matrix
element of the one-body density matrix
\cite{vit90}.
For shadow functions its non-normalized value reads

\begin{equation}\label{eq:pod}
\begin{aligned}
& \ \ {\cal P}_{od}(R,R',S,S') \propto \psi_a(R)\Theta_{SK}(R,S)\psi_s(S)\\
& \ \times \ \psi_a(R')\Theta_{SK}(R',S')\psi_s(S').
\end{aligned}
\end{equation}

After equilibration we just start binning values
proportional to  ${\cal P}_{od}$ as
a function of $r$.
In fact, to improve the statistical resolution of the algorithm,
we followed a further suggestion of Ceperley and Pollock\cite{cep87} and sampled 
instead

\begin{equation}\label{eq:pod}
\begin{aligned}
& \ \ {\cal P}_{od}(R,R',S,S') \propto \frac{1}{r^2n_a(r)}\psi_a(R)\Theta_{SK}(R,S)\psi_s(S)\\
& \ \times \ \psi_a(R')\Theta_{SK}(R',S')\psi_s(S').
\end{aligned}
\end{equation}
where $n_a(r)$ is a approximation
to the single-particle
density matrix that we take to be a Gaussian plus a constant. 
However the histogram we obtain is not normalized. Its normalization is
made by considering an
average of the first few values at small $r$ obtained by this method and
the previous one. This is possible, regardless if the kernel is symmetric
or not, because we choose $\rho_1(r) \rightarrow 1$ as $r \rightarrow 0$. 
This method is a complement to the first one we have described.

We have also estimated 
the pair distribution function of
atoms $g(r)$ defined as the probability of finding a pair of particles 
at a given separation $r$. The $g(r)$ is computed by taking the average

\begin{equation}
\label{eq:gor}
 g(r) = \frac{1}{N\rho}\left< \sum_{i<j}^N \delta(|{\bf r}_i - {\bf r}_j -
{\bf r}|) \right>
\end{equation}
with respect to ${\cal P}(R,S,S')$.
This quantity is estimated by updating by one
the bin of an histogram with bar width $\Delta r$  corresponding to
the relative distances between the atoms.
At the end of the simulation the histogram is normalized according the above
expression, taking into account how many
configurations were used.
Similar procedure was employed to compute the pair correlation function of
the shadow particles $S$ and $S'$ that were averaged to obtain the final
result.

\section{\label{sec:results} Results}

\subsection{Simulations}

Our simulations were carried out for systems with 108 particles for the liquid and 180 for the solid phases. 
In the liquid and solid phases the simulations started from a $fcc$ and an $hcp$ lattices, respectively.
Periodic boundary conditions were imposed.
Our runs consisted of $2.0 \times 10^5$ Monte
Carlo steps. Initially 8000 steps were discarded
to reach equilibrium.
Our Monte Carlo steps consisted of two
attempts to move the atoms followed by
three attempts to move each set of shadow coordinates.

The parameter space of the trial function
was exhaustively searched. 
The sets
that minimizes the energy expectation values as a function of the density
are presented in \tab{vp}. 
For shadow variables correlations of the
McMillan form at $\rho = 0.365\ \sigma^{-3}$ in the
liquid phase the best set of parameters is
given by
$\{b_a=1.13\ \sigma, C=5.1\ \sigma^{-2}, b_s = 1.29\ \sigma\}$, where $\sigma = 2.556 \angstrom$.

\begin{table}
\centering   
\setlength{\arrayrulewidth}{2\arrayrulewidth}  
\setlength{\belowcaptionskip}{10pt}  
\caption{\label{tab:vp}{Optimum variational parameters for the shadow wave function with
a symmetric kernel at the given densities
($\sigma =2.556\angstrom$).}}
\label{tab:vp}
\begin{tabular}{ccccc}
\hline
\hline
$\rho(\sigma^{-3})$ & $b_a(\sigma)$ & $C(\sigma^{-2})$ & $\beta(\text{K}^{-1})$ & $\alpha(\angstrom)$ \\
\hline
\multicolumn{5}{c}{Liquid} \\
$0.340$ & $1.12$ & $6.0$ & $0.058$ & $0.883$\\
$0.365$ & $1.12$ & $6.0$ & $0.060$ & $0.883$\\
$0.390$ & $1.12$ & $6.0$ & $0.060$ & $0.890$\\
$0.416$ & $1.10$ & $6.6$ & $0.074$ & $0.890$\\
$0.431$ & $1.10$ & $6.8$ & $0.068$ & $0.893$\\
\multicolumn{5}{c}{Solid} \\
$0.468$ & $1.07$ & $6.2$ & $0.100$ & $0.875$\\
$0.500$ & $1.09$ & $6.2$ & $0.070$ & $0.875$\\ 
$0.551$ & $1.09$ & $6.4$ & $0.060$ & $0.875$\\
$0.589$ & $1.11$ & $6.7$ & $0.060$ & $0.890$\\
\hline
\hline
\end{tabular}
\end{table}

\subsection{Variational Energies and equations of state}

\begin{figure}
\begin{center}
\includegraphics[scale=1.0]{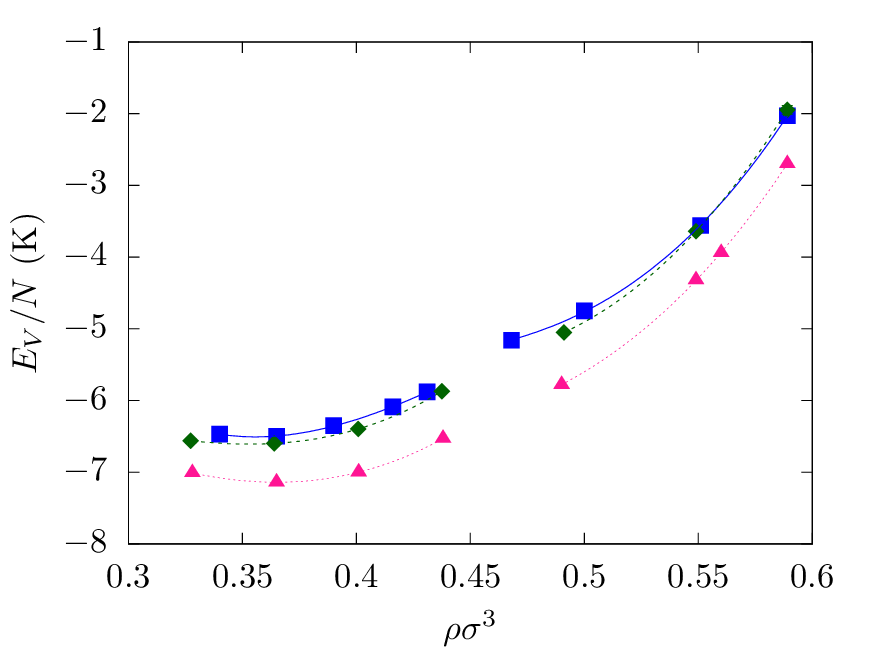}
\end{center}
\caption{\label{liqenergyfig} Ground-state energies
as a function of the density for the liquid and solid phases of a
system formed from 
$^4$He atoms. Blue squares represent our results and the solid blue line is a fit to the computed
values. The green diamonds and the dashed green line are the reported results in
Ref.~\onlinecite{mac94}.
In both phases pink triangles
stand for fits to values obtained by
neutron scattering experiments. The dotted pink line represent the fitting of these data.
In the liquid phase the fit is the one by
Aziz and Pathria\cite{azi73}.
For solid helium we have fitted the 
data from Woods et al.\cite{woo77}.
}
\end{figure}

Variational energies per atom obtained with the shadow function with a
symmetrical kernel for the liquid and solid phases are shown in \tab{ve}.
As expected \cite{mac94}, the energy at the experimental equilibrium density $\rho=0.365\ \sigma^{-3}$,
with correlation factors dependent on the rescaled Aziz
interactomic potential for the shadow variables, is lower than the one
obtained with correlation factors of the McMillan form.  We have estimated
the variational energy in this last case as $-5.91 \pm 0.02$ K,
\textsl{i.e.}, about
0.6 K higher than the one obtained with the rescaled Aziz potential.
Curves of the energy as a function of the density were fitted to the
estimated variational values using the expression of \eq{eos}. The results are
presented in Fig.~\ref{liqenergyfig}.
The fitted coefficients we have obtained are given in \tab{paramfit}.
The equilibrium density
from the fit,
$\rho_0 = 0.357\sigma^{-3}$,
agrees well with the experimental
value \cite{roa70}, $\rho_0 = 0.3649 \sigma^{-3}$

\begin{table} [!h] 
\setlength{\arrayrulewidth}{2\arrayrulewidth}  
\setlength{\belowcaptionskip}{10pt}  
\caption{
Variational energies per particle in units of K for a system formed from $^{4}$He at
the given densities.}
\label{tab:ve}
\begin{tabular}{ccc|ccc}
\hline
\hline
$\rho(\sigma^{-3})$ & $E_V/N$ & & & $\rho(\sigma^{-3})$ & $E_V/N$ \\
\hline
\multicolumn{2}{c}{Liquid} & & & \multicolumn{2}{c}{Solid} \\
$0.340$ & $-6.47 \pm 0.04$ & & & $0.468$ & $-5.16 \pm 0.03$ \\
$0.365$ & $-6.50 \pm 0.03$ & & & $0.500$ & $-4.75 \pm 0.06$ \\ 
$0.390$ & $-6.35 \pm 0.03$ & & & $0.551$ & $-3.56 \pm 0.03$ \\
$0.416$ & $-6.09 \pm 0.03$ & & & $0.589$ & $-2.03 \pm 0.01$ \\
$0.431$ & $-5.88 \pm 0.04$ & & & & \\
\hline
\hline
\end{tabular}
\end{table} 

\begin{table} 
\centering   
\setlength{\arrayrulewidth}{2\arrayrulewidth}  
\setlength{\belowcaptionskip}{10pt}  
\caption{\label{tab:paramfit}{Adjusted coefficients $A$, $B$ and $C$ in units of K
and $\rho_0$ in units of $\sigma^{-3}$,
of \eq{eos}, the polynomial fit
to the variational energies.}}
\begin{tabular}{lcccc}
\hline
\hline
   & $A$  & $B$  & $C$  & $\rho_0$ \\
\hline
 Liquid & -6.51 & 17.70 & -15.57 & 0.357\\
 Solid & -5.42  & -0.01 & 19.45 & 0.378\\
\hline
\hline
\end{tabular}
\end{table}

In the solid phase we can see that as the density increases,
our energy becomes marginally lower than the results of MacFarland
\textsl{et al.} \cite{mac94}.
To some extent a similar behavior can also be seen in the liquid state
where as the density increases we see our variational energies approaching
those of Ref.~\onlinecite{mac94}.
Since in our trial function the sampling of 
exchange between atoms is more efficiently done,
these results suggest that 
the importance of exchange  increases with the density.

\subsection{Melting-Freezing Transition}

The melting and freezing densities are easily determined through the EOS
of the liquid and solid phases using the double tangent Maxwell
construction, \eq{mft}.
The value we have estimated for the 
freezing density is $\rho_f = 0.457\ \sigma^{-3}$. It can be
compared with the experimental value of 0.431 $\sigma^{-3}$. For the
melting transition our calculation gave 
$\rho_m=0.495\ \sigma^{-3}$ and experiment 0.468 $\sigma^{-3}$.
Although our melting transition density is about of the same quality
obtained with a shadow function with
optimized two-body correlations between atoms \cite{mor98},
our freezing density it not so good. 

\subsection{Radial distribution functions}

\begin{figure} 
\begin{center}
\includegraphics[scale=1.0]{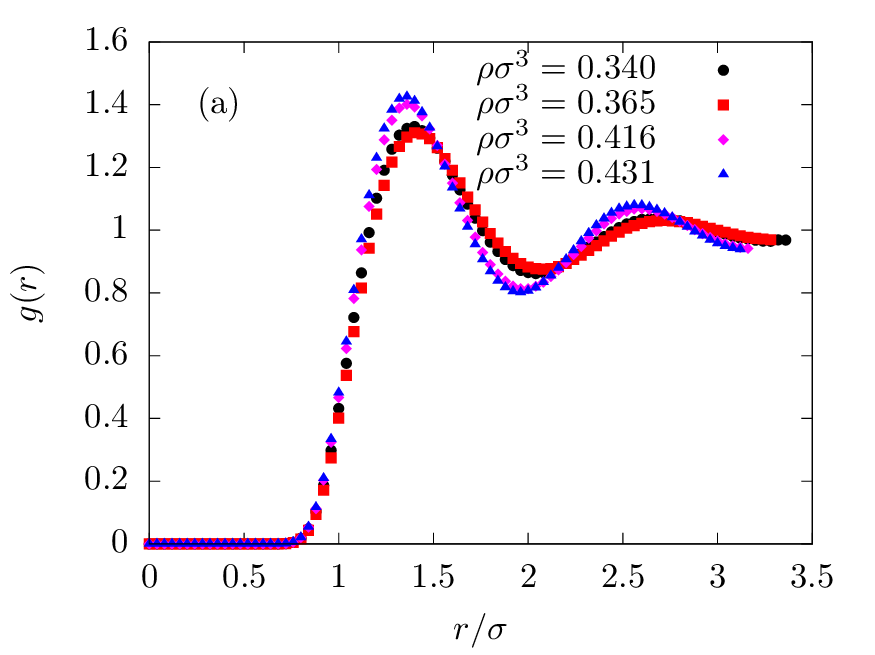}
\includegraphics[scale=1.0]{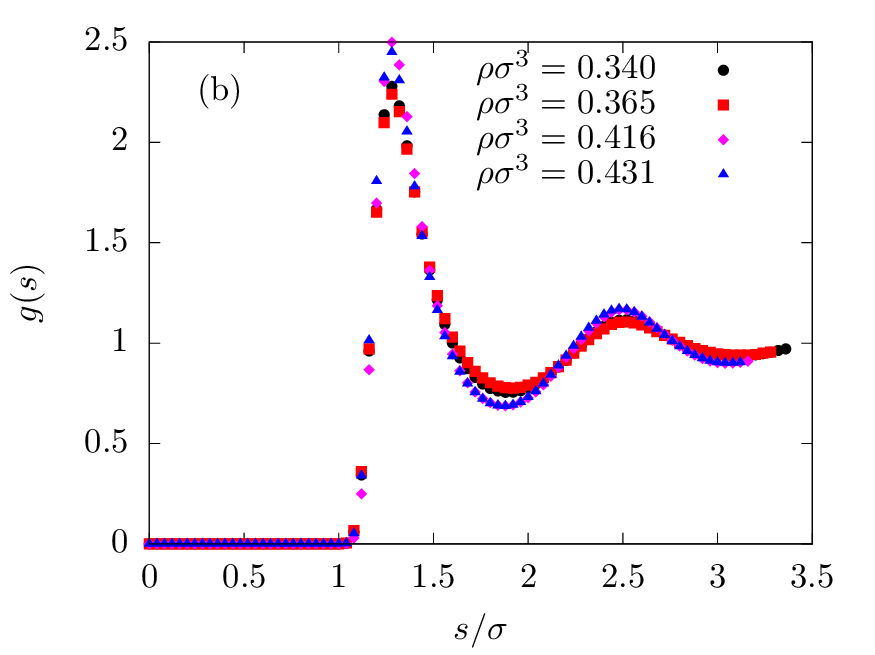}
\end{center}
\caption{\label{fig:liqg(r)}{ Radial distribution of atoms $g(r)$ (a) and
auxiliary variables $g(s)$ (b) at four densities in the liquid phase
of the $^4$He system.  As expected the system turns out to have more structure
as the density increases. Shadow particles have peaks much more intense.}}
\end{figure}

\begin{figure} 
\begin{center}
\includegraphics[scale=1.0]{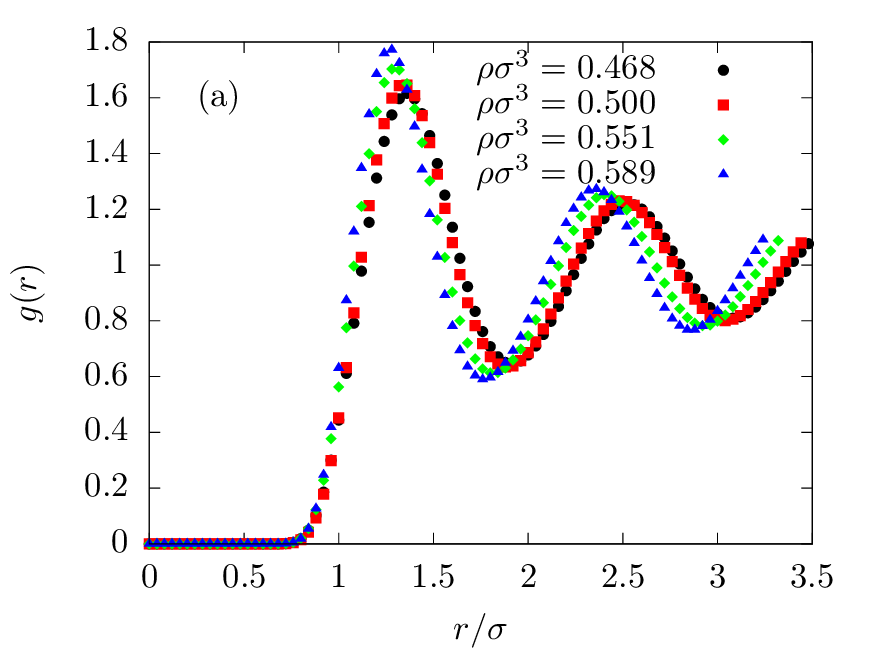}
\includegraphics[scale=1.0]{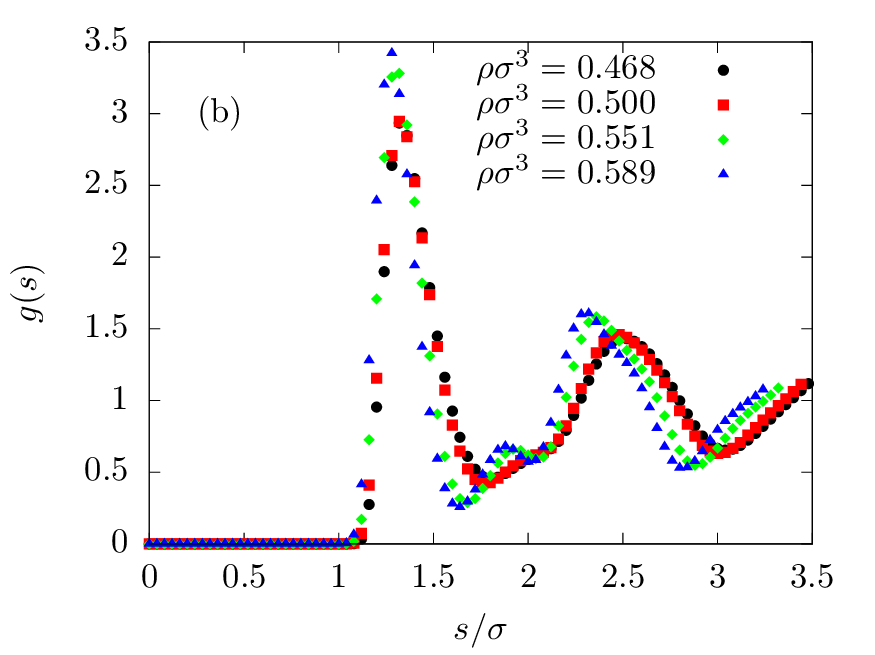}
\end{center}
\caption{\label{fig:solg(r)}{ Radial distribution of atoms $g(r)$ (a) and auxiliary
variables $g(s)$ (b) at four densities of the $^4$He in the solid
phase.
}} \end{figure}

The radial distribution functions $g(r)$ computed at four densities in the liquid phase is shown in \fig{liqg(r)}.
 The figure on the left is the radial distribution function of the atoms of
The shadow particles, that model the center of mass of polymers of the Feynman
path-integral in imaginary time
reflect somewhat the  {\em more classical} behavior of this particles.
This is also the behavior we see for the shadows in the crystal case displayed
in \fig{solg(r)}. It is visible the formation of
a small shoulder before the second peak
typically seen in the crystallization process of classical fluids. 
The radial distribution of the atoms in the solid phase displayed at the
same figure show more structure than the liquid phase. However it is much
less pronounced than in classical solids.

\begin{figure} 
\begin{center}
\includegraphics[scale=1.0]{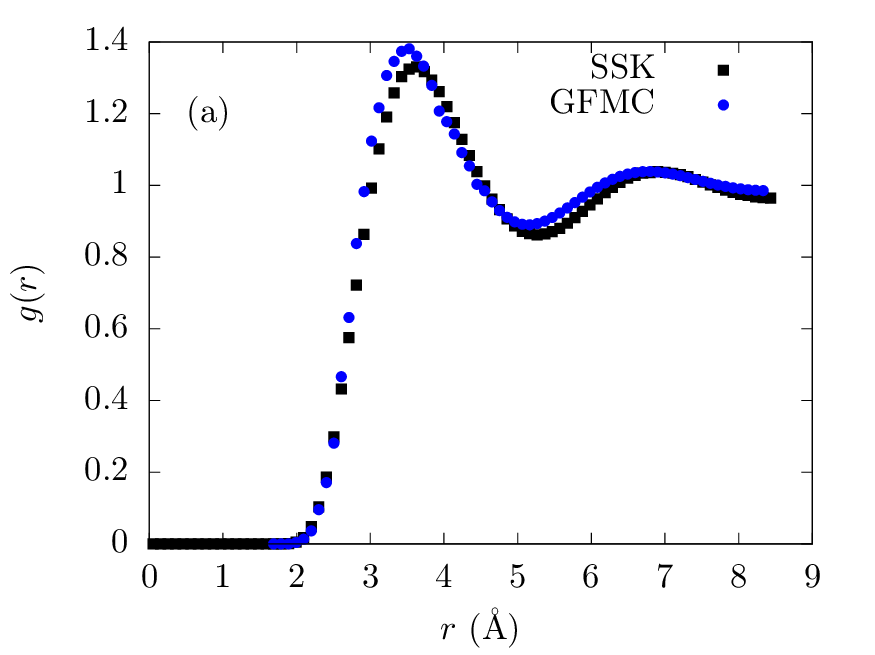}
\includegraphics[scale=1.0]{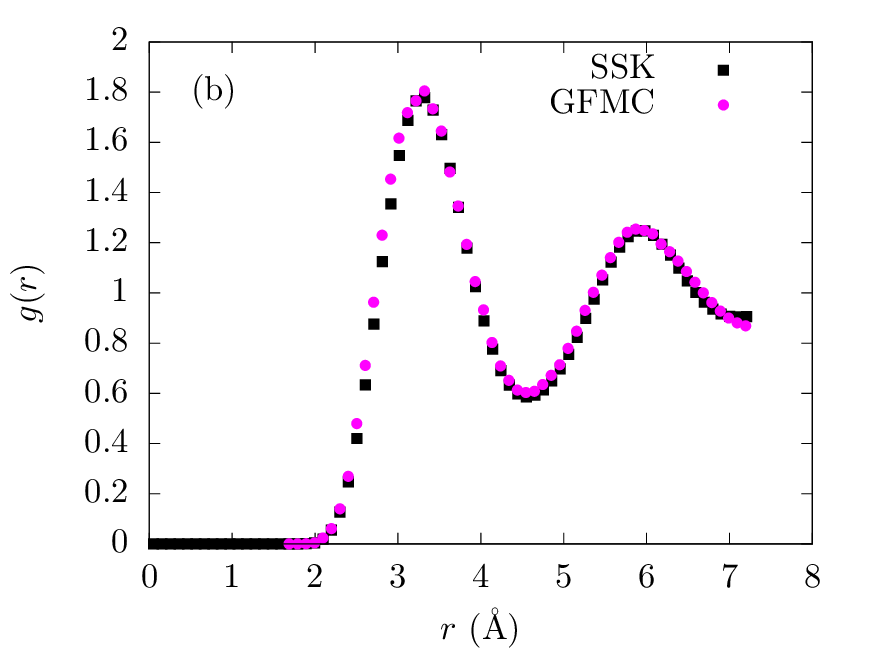}
\end{center}
\caption{\label{fig:liqcompg(r)}{ Comparison of $g(r)$ determined with the
$\Psi_{SSK}$ trial function (black squares) with
GFMC results (blue and pink circles) from Kalos \textsl{et al}. \cite{kal81} for
$^4$He systems.
For the liquid phase (a) the calculations are at the experimental equilibrium
density $\rho = 0.365\sigma^{-3}$
and for the crystal phase (b)
at $\rho = 0.589\sigma^{-3}$.
}}
\end{figure}

It is also interesting to compare our results
for the radial function of atoms
to those obtained by the
GFMC method\cite{kal81} that gives essentially the exact results.
This is made in \fig{liqcompg(r)}.
It is worth to mention that the
inter-atomic potential we use \cite{azi95}
is a more recent one. It does not include in an effective way three-body
contributions like the one \cite{azi79}
employed in the GFMC calculation.
The agreement of the results in the liquid
phase are very good despite the difference of the
potentials used in the two calculations.
At the solid phase at the density $\rho = 0.589\sigma^{-3}$ there is a remarkable  agreement
as we can see in \fig{liqcompg(r)}(b). It is at this density that our trial function
outperform shadow functions without an explicit symmetric kernel.

\subsection{Condensate fraction}

\begin{figure} 
\begin{center}
\includegraphics[scale=1.0]{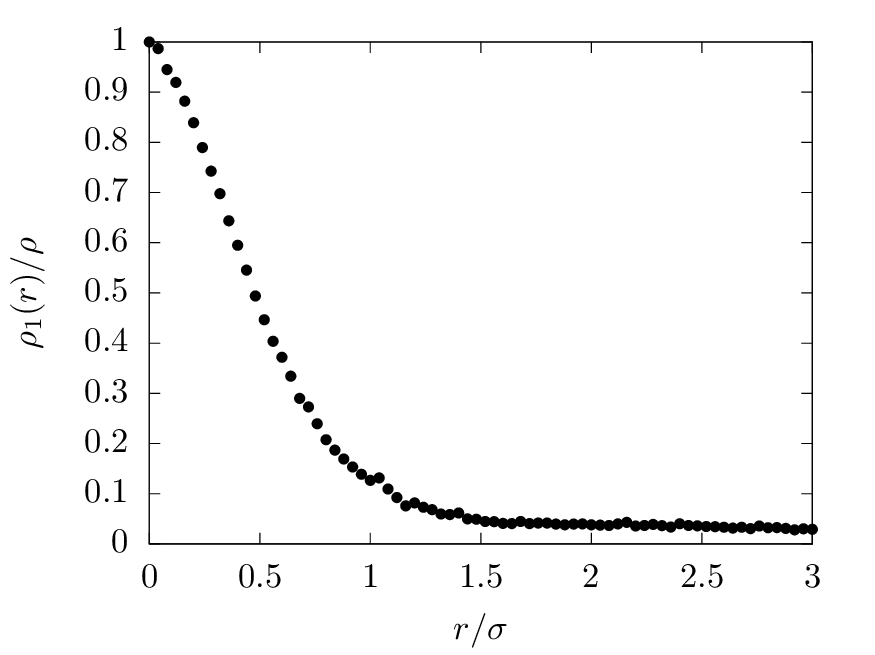}
\end{center}
\caption{\label{fig:obdm}{ The one-body density matrix estimated using
\eq{density_matrix}
at the experimental
equilibrium density $\rho = 0.365\sigma^{-3}$.}}
\end{figure}

\begin{figure} 
\begin{center}
\includegraphics[scale=1.0]{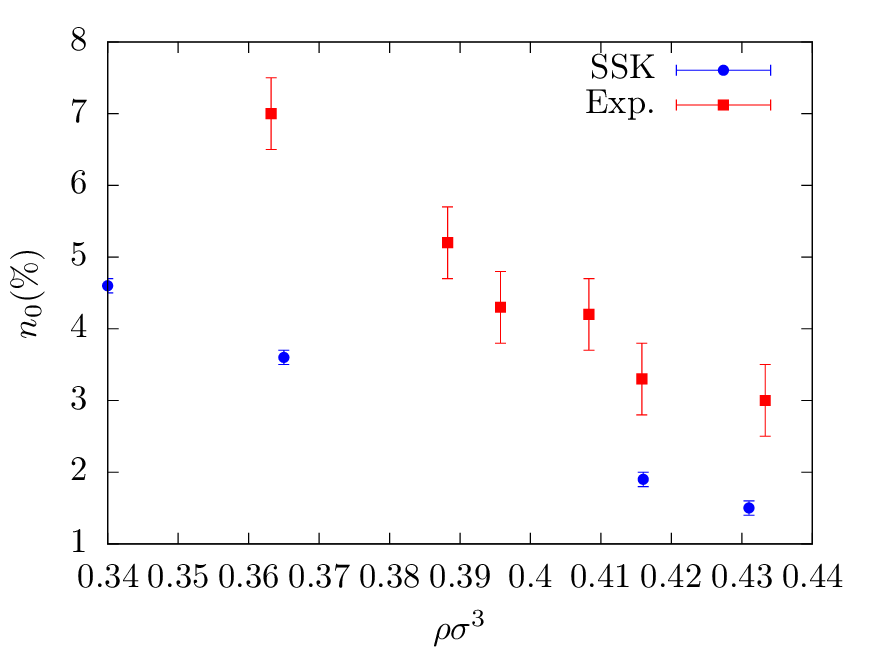}
\end{center}
\caption{\label{fig:f0comp}{ The condensate fraction
estimated trough the sampling of ${\cal P}(R,S,S')$ at
four densities at the liquid phase compared with experimental
results extracted from Ref.~\onlinecite{gly11}.}}
\end{figure}

The single particle momentum distribution $n(k)$ characterizes the
extent a system formed from $^4$He have a behavior
that deviates from classical physics. The
strong quantum effects present in this system
does not allow a description in terms of
the Maxwell-Boltzmann
distribution, typical of classical systems.
A related quantity,
the off-diagonal matrix element of the one-body density matrix,
$\rho_1(r)$ can be calculated
straightforward
through \eq{density_matrix}.
At the equilibrium experimental density
$\rho = 0.365\sigma^{-3}$,
our results are shown in \fig{obdm} from where
we have extracted 
$n_0 = (3.57 \pm 0.07)\%$
for the condensate fraction.
We have computed this quantity at other densities as well, the results
together with experimental measures \cite{gly11rc} are
displayed in \fig{f0comp}.

\begin{figure} 
\begin{center}
\includegraphics[scale=1.0]{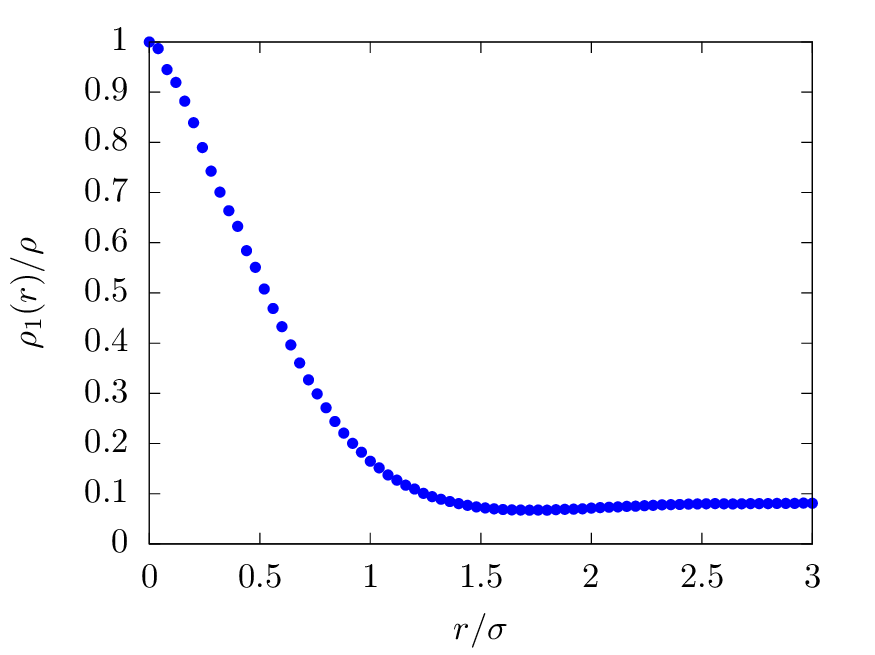}
\end{center}
\caption{\label{fig:normrho1}{ The normalized one body density function obtained
by sampling \eq{pod}.
The associated condensate fraction is $(8.17 \pm 0.03)$ at the experimental equilibrium density.}}
\end{figure}

At the equilibrium density the theoretical and experimental values of the
condensate fraction differ by
a factor of about 2. This fact lead us to consider
the question of how efficient the sampling
of \eq{probability} can be for the
condensate fraction estimation.
For this reason we have also considered sampling the probability
distribution function of \eq{pod} 
to compute $\rho_1(r)$.
Its normalization factor can be obtained from \fig{obdm}
by considering
small values of $r$.
The normalized result of $\rho_1(r)$ determined in this
way is presented in
\fig{normrho1}. From this calculation
the associated condensate fraction is equal to
$n_0 = (8.17 \pm 0.03)\%$. A value in much better agreement
with the experimental data\cite{gly11},
$(7.25 \pm 0.75)\%$.

Although we are in front the apparent puzzle of having two different
values for the estimation of a given property
(the condensate fraction),
it is possible to explain these results.
The shadow particles create about themselves through the Jastrow factor $\psi_s$
a much larger exclusion
volume than the one of the atoms. This situation produces a jam in their
moves. For feasible computational times the shadow particles are not able to effectively
explore the phase space available to them. 
Certainly, if we could wait for the shadow
moves through their jam, both ways of computing the single particle momentum
distribution would agree.
The result of \fig{obdm} are due only to the symmetrization we have
introduced in our trial function but does not take into account, or maybe
takes only partially, the possibility of the diffusion of the shadow
particles in their configuration space.

For completeness and to be careful with our results
we have also investigated how finite
size effects of the simulation box might
affect the condensate
fraction calculations
when we sample the
probability density function of \eq{pod}.
We carried out simulations for systems of 32, 64 and
108 bodies at the experimental equilibrium density $\rho =
0.365\sigma^{-3}$. The results are shown in \tab{boxsize}.
From our results it is possible to say that with 64 bodies 
finite size effects most probably are negligible. Nevertheless all the
reported results were obtained considering $N=108$ bodies.

\begin{table} 
\centering   
\setlength{\arrayrulewidth}{2\arrayrulewidth}  
\setlength{\belowcaptionskip}{10pt}  
\caption{\label{tab:boxsize}{The condensate fraction $n_0$ estimated for
systems of 32, 64 and 108 bodies at
the experimental equilibrium density ($\rho\sigma^3 = 0.365$).
}}
\begin{tabular}{cccc}
\hline
\hline
$N$ & 32 & 64 & 108 \\
\hline
{$n_0$ ($\%$)\; }  & $9.2 \pm 0.4$\; & $8.3 \pm 0.1$\;  & $8.17 \pm 0.03$\\
\hline
\hline
\end{tabular}
\end{table} 

\section{\label{sec:comments} Final comments}

The shadow wave function is a powerful tool to investigate quantum liquids
and solids formed from helium atoms. Since its inception \cite{vit88},
steady progress has been made. First an attractive pseudo-potential and
optimized two-body correlations were introduced \cite{mac94}. Later it was
extended to treat the fermionic system made from $^3$He atoms
\cite{ped96}. More recently \cite{dan09} a much more sophisticated
approach to the last problem was introduced where the antisymmetric character 
of the wave function was introduced trough the auxiliary variables themselves. 
In this work we have
modifyied the atom-shadow coupling by introducing an explicitly 
symmetric kernel and analyzed its consequences.
In a formal way this approach does change basic properties of this class
of trial functions like translational and its symmetrical character.
However it was possible to demonstrate that exchange correlations becomes
more important as the density increases. We have also shown the need of
considering in an explicit way the jam created by the shadow particles in
the liquid phase for computing the condensate. As we increase the density in
the solid phase we are able to improve the variational energy with respect
to a kernel not symmetric.
The shadow function has implicit correlations up to the number of
particles considered in the system. We believe that attempts to optimize
these correlations are important because they might help to uncover yet
unknown properties of the systems formed from helium atoms.

\begin{acknowledgements}
The authors acknowledge financial support from the Brazilian agencies
\textsc{fapesp} and \textsc{cnp}q. Part of the computations
were performed at
the \textsc{cenapad} high-performance computing facility at Universidade
Estadual de Campinas.
\end{acknowledgements}

\bibliographystyle{apsrev4-1}
\clearpage

%

\end{document}